\documentstyle[12pt]{article}

\begin{document}

\sloppy

\begin{flushright}{UT-681\\ May  '94}\end{flushright}
\vskip 1.5 truecm

\centerline{\large{\bf Natural explanation for discrete R-symmetry }}
\centerline{\large{\bf in successful inflation in N=1 supergravity }}
\vskip .75 truecm
\centerline{\bf Tomohiro Matsuda}
\vskip .4 truecm
\centerline {\it Department of Physics,University of Tokyo}
\centerline {\it Bunkyo-ku,Tokyo 113,Japan}
\vskip 1. truecm

\makeatletter
\@addtoreset{equation}{section}
\def\theequation{\thesection.\arabic{equation}}
\makeatother

\vskip 1. truecm

\begin{abstract}

Recently it was shown that discrete R-invariance in superpotential
can lead to a successful flat potential in inflation.
We suggest that this discrete R-symmetry arises from an underlying
supersymmetric gauge theory, which gives rise to a scalar inflaton as a
composite field.
 This discrete R-symmetry is common to the dynamical breaking
scenarios in global supersymmetry  studied several years ago.
 Some extension
 of the R-invariance method is also shown.

\end{abstract}
\newpage

\section{Introduction}
Although supersymmetric theories seem very attractive for grand
unification[1], there are some problems. It seems that low energy theories
are well explained by N=1 supergravity theories, but there are
potential cosmological problems. One of them is the difficulty to
construct the inflation scenarios of the universe.
\par
Many models are proposed to solve these problems. Recently Kumekawa et.al[2].
have proposed a successful inflation model imposing $Z_{n}$ symmetry on
superpotential. This model has a very flat potential for inflation, and
reheating temperature $T_{R}$, and gravitino mass
$m_{\frac{3}{2}}$ are reasonable
 if we choose $Z_{4}$.
\par
Using a composite field for the inflaton field,
we suggest that the origin of the $Z_{n}$ symmetric inflation
may be explained naturally by field condensation.
We also consider what happens if we introduce another
matter field for the hidden sector.

\section{Review of discrete R-symmetric model}
In this section we briefly review the idea of ref.[2].
We define a discrete $Z_{n}$ R-transformation on the inflaton field
$\phi(x,\theta)$ as
\begin{equation}
\phi(x,\theta)\rightarrow{e^{-i{\alpha}}}\phi(x,e^{\frac{i}{2}\alpha}
\theta)\ \  ,\ \ \alpha=\frac{2\pi{k}}{n}\ \ (k=0,\pm1,\pm2,...).
\end{equation}

General form for a K\"{a}hler potential $K(\phi,\phi^{*})$ and a
superpotential $W(\phi)$ for the inflaton are given by
\begin{eqnarray}
K(\phi,\phi^{*})&=&\sum^{\infty}_{m=1}a_{m}(\phi\phi^{*})^{m}\\
W(\phi)&=&\phi\sum^{\infty}_{l=0}b_{l}\phi^{l\,n}.
\end{eqnarray}
{}From now on, we take the minimum K\"{a}hler
 potential ($a_{1}=1, a_{i}=0\ \
 for \ \ i\ne1$).
 To see the shape of the potential near the origin, we write
superpotential as

\begin{equation}
W(\phi)=\left(\frac{\lambda}{v^{n-2}}\right)\left(v^{n}\phi-
\frac{1}{n+1}\phi^{n+1}\right)+\cdots.
\end{equation}
Here $\lambda$ is a coupling constant and $v$ is a scale factor.
A scalar potential can be derived using the relation

\begin{equation}
V(\phi)=exp\left(\frac{K(\phi,\phi^{*})}{M^{2}}\right)
\left\{(K^{-1})^{\phi^{*}}_{\phi}(D_{\phi}W)(D_{\phi}W)^{*}-
\frac{3|W|^{2}}{M^{2}}\right\}
\end{equation}
and taking the approximation $\phi\sim0$, we get

\begin{equation}
V(\phi)\simeq\left(\frac{\lambda}{v^{n-2}}\right)^{2}\left\{v^{2n}
\left(\frac{|\phi|^{2}}{M^{2}}\right)^{2}-v^{n}(\phi^{n}+\phi^{*n})
\right\}.
\end{equation}
Where $M$ is the gravitational mass $M=M_{pl}/\sqrt{8\pi}\simeq2.4\times
10^{18}GeV$.
When $n=2$, this potential does not have a flat potential.
For $n=3$ and more, we can neglect the
$\left(\frac{|\phi|^{2}}{M^{2}}\right)^{2}$ term as ref.[2].
Setting the phase of $\phi$ to vanish, we get

\begin{equation}
V(\varphi)\simeq\tilde{\lambda}\tilde{v}^{4}\left\{1-2(
\frac{\varphi}{\tilde{v}})^{n}\right\}.
\end{equation}
Here we set

\begin{eqnarray}
\varphi&\equiv&\sqrt{2}Re\phi\\
\tilde{\lambda}&\equiv&\frac{1}{2}\lambda\\
\tilde{v}&\equiv&\sqrt{2}v.
\end{eqnarray}
This potential is very flat near the origin.
Using this potential, we can calculate Hubble constant $H$, e-folding factor
 $N$, density fluctuation
$\frac{\delta\rho}{\rho}$, gravitino mass $m_{\frac{3}{2}}$, and inflaton
mass $m_{\phi}$,
 etc. In this case, $\Lambda_{cos}$ should be fine tuned introducing a D-term.

\par
Naively, decay width of the inflaton field $\Gamma_{\phi}$ is so small
that reheating is not enough. But this problem can be avoided introducing
 another singlet chiral super multiplet with a half $Z_{n}$ charge [2].
\par
This model can explain successful inflation, particularly for $Z_{4}$ case.

\section{Natural explanation for $Z_{n}$ symmetry}
\par
Let us consider supersymmetric Yang Mills (SYM) case,
 with SU(N) gauge symmetry. In this section, we follow the notations of
ref.[3].
\begin{equation}
L_{SYM}=\frac{1}{4}\int{d}^{2}{\theta}\ W^{a}W^{a}+\frac{1}{4}{\int}d^{2}
\overline{\theta}\ \overline{W}^{a}\overline{W}^{a}
\end{equation}
In this Lagrangian, there exists global R-symmetry $U(1)_{R}$ [3]
for the gaugino field $\lambda$ as
\begin{equation}
\lambda\rightarrow{e}^{i\alpha}\lambda
\end{equation}
This symmetry has anomaly, and is broken at the quantum level.
But its  subgroup $Z_{2N} (\alpha\equiv\frac{2\pi{k}}{2N},k=1,...,2N)$
 still unbroken[3].
\par
Considering gaugino condensation, we introduce a composite field.
\begin{equation}
S=\frac{g^{2}}{32\pi^{2}}W^{a}W^{a},
\end{equation}
with $<S>\ne0$. This field has desired discrete R-symmetry $Z_{N}$.
\par
Taking this composite field as inflaton field, we can get a flat
potential as is described above.
\par
Here we should mention the anomaly matching condition relating to the
global R-symmetry(3.2).
If we consider this condition, we get a logarithmic potential [2].
\begin{equation}
W(S){\sim}SlogS^{N}.
\end{equation}
This superpotential leads to the scalar potential which has a logarithmic
singularity at the origin $S=0$.
This singularity stems from the fact that near the origin $S\simeq0$,
the anomaly cannot be simply computed in terms of the field $S$ but should
be calculated by taking into account the dynamical degrees of freedom
contained in the theory before the condensation, such as $\lambda$.
Therefore, denoting the typical energy scale of the gaugino condensation
as $\Lambda$, when $S\sim\Lambda$, we cannot describe the theory in terms of
the composite field $S$ alone, and the anomaly matching condition for the
superpotential breaks down.
Even in such regions(i.e.,$S\sim\Lambda$),
 when the condensation is almost completed, the behavior
of the order parameter $S$ will be still described by the effective potential
which is regular near the origin.
This potential should be $Z_{N}$ symmetric since
$Z_{N}$ symmetry always exists regardless whether we implement the anomaly
matching condition in the scalar freedom or not,
and the potential,which is analogous to the free energy,
 should not have singularity at the origin.
This assumption seems reasonable except for the choice of
applicable domain of the
effective scalar potential,
which should be assumed by hand for the successful inflational scenario.
 For example, in order to have successful
inflation, potential should satisfy above conditions ($Z_{N}$ symmetry and
no singularity near the origin) \underline{at least} in the region
\begin{equation}
v(\frac{v}{M})^{4/(N-4)} < S < (\frac{3}{2N(N-1)}\frac{\tilde{v}^{N}}
{M^{2}})^{1/(N-2)}.
\end{equation}
The lower bound in (3.5) arises from the initial fluctuation and the
upper bound is required for sufficient inflation [2].
When S is smaller than (3.5), the fermionic freedom $\lambda$
would dominate, and when $S$ is much
larger than (3.5),
the potential would be well described by (3.4).

\par
Assuming that the flatness can be satisfied on the basis of a physical
picture suggested above, the supersymmetry breaking
required for phenomenological analysis is not quite
satisfied in these simple models because the potential far from the origin is
 different from that in ref.[2]. The supersymmetry breaking by gaugino
condensation have been studied
in detail in[5](see also [6] and [7]),
 and we can apply this $Z_{n}$ mechanism to
 more complicated
models, but this is beyond the scope of this paper. This problem
 will be studied
elsewhere.
\par
Related to this problem, the constraints to the parameter are  different
from these in ref.[2].
In our model, $ v $ is not the vacuum expectation value of $S $ at the
potential minimum.
Constraints arising from successful inflation to v are the same but those
 from supersymmetry
breaking are very different from ref.[2].

We now comment on some related problems.
We can add matter fields.
For simplicity, here we consider supersymmetric QCD (SQCD)[3].
Lagrangian is given by

\begin {eqnarray}
L_{SQCD}&=&\frac{1}{4}\int{d}^{2}{\theta}\ W^{a}W^{a}+\frac{1}{4}{\int}d^{2}
\overline{\theta}\ \overline{W}^{a}\overline{W}^{a}\nonumber\\
&&+{\int}d^{4}\theta[\Phi^{+i}e^{2gV^{a}T^{a}[N]}\Phi_{i}+\tilde{\Phi}^{+i}
e^{2gV^{a}T^{a}[\tilde{N}]}\tilde{\Phi}_{i}]\nonumber\\
&&+{\int}d^{2}\theta{m}^{i}_{j}\Phi_{i}\tilde{\Phi}^{j}+
{\int}d^{2}\overline{\theta}{m}^{i}_{j}\overline{\Phi_{i}}
\overline{\tilde{\Phi}}^{j}.
\end{eqnarray}
In this case,
global R-symmetry becomes (in terms of composite fields)
\begin{eqnarray}
S&\rightarrow&e^{-2i\alpha}S(x,e^{i\alpha}\theta)\\
T&\rightarrow&e^{-2i\alpha}T(x,e^{i\alpha}\theta).
\end{eqnarray}
This R-symmetry is also broken by anomaly.
Here we used the definition
\begin{equation}
T^{j}_{i}\equiv\Phi_{i}\tilde{\Phi}^{j}.
\end{equation}
There is another global symmetry, called X-symmetry.
\begin{eqnarray}
S&\rightarrow&e^{-2iM\alpha}S(x,e^{iM\alpha}\theta)\\
T&\rightarrow&e^{2i(N-M)\alpha}T(x,e^{iM\alpha}\theta)
\end{eqnarray}
Here $M$ means number of flavor. This X-symmetry is a combination of
axial and R-symmetry, and anomaly-free.
Now we can construct superpotential for SQCD.
For example,
\begin{equation}
W(S,T)=S\sum^{\infty}_{l=0}c_{l}(S^{N-M}|T|)^{l}
\end{equation}
Notice that this potential has large Yukawa couplings.
This means, even though we could make a successful flat potential for
inflation, almost all energy runs away to the hidden sector.
The reheating of hidden sector results in dark matter or nothing. Of
course, reheating in the observable sector is extremely suppressed.

\section{Conclusion}
We have suggested that the gaugino condensation in the hidden sector can
lead to the successful inflation, if one assumes that the behavior of the
order
parameter $S$ from the gaugino condensation is controlled by the effective
potential(which is singularity-free)
 in the vicinity of the condensation energy scale.
This is because the basic underlying theory respects the $Z_{n}$ symmetry as is
required for the successful inflation in ref.[2].
We also analyzed some extension to include mater fields.
\par
There are some problems to be resolved.
To consider inflation, we are interested in the shape of the potential
near the origin, which in turn means that we should consider the
critical behavior of
the condensation.
Validity of our basic picture described in this note depends on the detailed
dynamics of the condensation, and it's analysis is beyond the scope
of our discussion.
\par
Our analysis in this paper should be regarded as a proposal of a possible
physical picture behind the $Z_{n}$ symmetry.
\subsection*{Acknowledgments}
We thank K.Fujikawa, T.Hotta and A.Yamada for helpful discussions.


\begin{thebibliography}{1}

\bibitem{1}
E.Cremmer, S.Ferrara, L.Girardello, and A.van Proyen,
 Nucl.Phys.{\bf B212}(1983)413

\bibitem{2}
K.kumekawa, T.Moroi, and T.Yanagida, preprint TU-458(1994)
\bibitem{3}
D.Amati, K.Konishi, Y.Meurice, G.Rossi and G.Veneziano,
 Phys.Rep.{ \bf 162 }(1988)162

\bibitem{4}
J.Wess and J.Bagger, $ Supersymmetry \  and \  Supergravity $ ,\\
(Princeton University Press, 1992)

\bibitem{5}
J.P.Derendinger, L.E.Iba$\tilde{n}$ez, and H.P.Nills,
Phys.Lett.{\bf B155}(1985)65\\
M.Dine, R.Rohm, N.Seiberg, and E.Witten,
Phys.Lett.{\bf B156}(1985)55\\
T.R.Taylor, Phys.Lett.{\bf B252}(1990)59\\
B.de Carlos, J.A.Casas, F.Quevedo and E.Roulet,
 Phys.Lett.{\bf B318}(1993)447

\bibitem{6}
Ewan D.Stewart 'INFLATION, SUPERGRAVITY AND SUPERSTRINGS'\\
 hep-ph/9405389

\bibitem{7}
T.Banks, D.B.Kaplan, A.E.Nelson, Phys.Rev.{\bf D49}(1994)779
\end{thebibliography}
\end{document}